\documentstyle[preprint,aps]{revtex}
\begin{document}
\preprint{UM-P-94/133; RCHEP-94/37}
\draft
\title{MAGNETIC MONOPOLE AND THE FINITE PHOTON MASS: ARE THEY COMPATIBLE?}
\author{A.Yu.Ignatiev\cite{byline1} and
G.C.Joshi\cite{byline2}}
\address{Research Centre for High Energy Physics, School of
Physics, University of Melbourne, Parkville, 3052, Victoria,
Australia}
\date{December 19, 1994}
\maketitle
\begin{abstract}
We analyze the role played by the gauge invariance for the existence of Dirac
monopole. To this end, we consider the electrodynamics  with massive photon and
ask if the magnetic charge can be introduced there.
We show that the derivation of the Dirac quantization condition based on the
angular momentum algebra cannot be generalized to the case of massive
electrodynamics. Possible implications of this result are briefly discussed.
\end{abstract}
\pacs{}
The question of whether magnetic monopoles exist in Nature has
been widely considered over the last six decades, beginning with the
pioneering work by Dirac \cite{d}. The interest in the theory of the monopole
\cite{rev} has not been diminished by the continuos failure to discover it in
the experiment
\cite{pd}.

In all these works, the gauge invariance of the Maxwell electrodynamics
has been assumed.  Meanwhile, the question was raised by de Broglie,
Schr\"{o}dinger and others if the Maxwell equation can be generalized
to include a new fundamental parameter with the dimension of length, $L$
\cite{gn}.
One consequence of these generalized Maxwell equations (called the Proca
equations) is that the electric field of a point like electric charge
would extend only over distances $r\sim L$ instead of infinite range of the
ordinary Coulomb force.  More exactly, the electric field would take the form
$E\sim\frac{1}{r^2}e^{-\frac{r}{L}}$.  This form is of course reminiscent
of the familiar Yukawa potential.  Indeed, it can be shown that in the quantum
language the introduction of the length $L$ means the existence of the photon
mass $m_{\gamma}=\frac{\hbar}{cL}$.

But in this paper we shall deal only with static classical fields so
we will use the terms "photon mass" or "massive electrodynamics" synonymously
with "finite range electrodynamics".

In this work, our question is: can we marry magnetic monopoles with nonzero
photon mass?  In other words, is there a consistent generalization of finite
range
Maxwell equations which would describe both electric and magnetic charges?

There are at least two reasons why this study is important. First, we need to
understand better the role of gauge invariance in the derivation of the Dirac
quantization condition. It is obvious that the condition of gauge invariance
figures prominently in a number of derivations proposed so far (such as the
original Dirac's proof \cite{d} or the Wu-Yang formulation \cite{wy}). However,
several alternative methods have been developed to derive the quantization
condition without making use of gauge invariance, single-valuedness of the
wave-function and the "veto" postulate forbidding charge particles from
crossing the string. Instead of gauge invariance, most of these works were
based on more general group-theoretical methods involving the rotational
invariance and the angular momentum quantization \cite{lwp,h,p,g}. Yet it is
not clear whether these methods can be generalized to a situation where the
gauge invariance is absent.

The second reason is the long-standing puzzle of the electric charge
quantization: why the electric charge of any elementary particle is a multiple
of that of the $d$ quark? Despite the fundamental character and apparent
simplicity of this phenomenon, we still lack a complete understanding of it.
Historically, the possible existence of the Dirac monopole was the the first
explanation of the charge quantization. (In fact, the whole Dirac's work on
monopoles was motivated by and grew out from his attempts to explain why the
charge is quantized.) The massive electrodynamics is the simplest extension of
the standard Maxwell theory; it is interesting to know whether the Dirac
argument can be extended to it or not.

Note that this paper should not be taken as advertising the theory with
non-zero photon mass; yet we feel that this theory is worth studying further,
whether one considers it aesthetically appealing or not.

Without going into detail here, we recall only the central, well-established
fact: massive electrodynamics is a perfectly consistent quantum field theory
\cite{gn}. In all respects (quantization, renormalizability, the electric
charge conservation and so on) it enjoys the same status as the standard QED.
In fact, with  massive electrodynamics the theorist's life  is sometimes
easier: for instance, it allows for a manifestly covariant quantization without
the need to introduce an indefinite metric. Also, the infrared behaviour of
massive QED is much simpler than that of standard  QED.

On the experimental side, the possible magnitude of the photon mass is severely
bounded: from the consideration of the galactic magnetic field $m_{\gamma} \alt
10^{-36}$ GeV \cite{c}, while other (more reliable but weaker) limits are
typically ten to fifteen orders of magnitude more relaxed \cite{pd}.

At first sight, there is nothing wrong with the co-existence of magnetic
monopoles and massive photons.  One would expect the only difference: the
"Coulomb" magnetic field of the monopole $H\sim\frac{1}{r^2}$ has to be
substituted by the "Yukawa" field: $H\sim\frac{1}{r^2}e^{-mr}$.

We will show that this simple picture is not the case:  the finite photon
mass and existence of the magnetic charge are incompatible with each other.

We start by writing down the Proca equation describing electrodynamics with
finite range (or equivalently with a non-zero photon mass $m$)\footnote{We use
the Heaviside system
of units throughout this paper and also put $\hbar =1, c=1$.}:

\begin{eqnarray}
   & \partial^{\mu }F_{\mu \nu } = J_\nu  - m^2A_{\nu} &\\
   & \partial^\mu  \tilde{F}_{\mu \nu } = 0&\\
   & \partial^\mu  A_\mu  = 0&\\
   &  F_{\mu \nu } = \partial_\mu A_\nu  - \partial_\nu A_\mu. &
\end{eqnarray}
The dual pseudo--tensor $\tilde{F}_{\mu \nu }$ is defined as usual via
$
       \tilde{F}_{\mu \nu } = \frac{1}{2}\varepsilon_{\mu \nu
\alpha\beta}F^{\alpha\beta}$;
$J_\nu $ is the (electric) current density.  Next, the Maxwell equations
generalized to
include magnetic charge are:

\begin{eqnarray}
   & \partial^\mu F_{\mu \nu } = J_\nu & \\
   & \partial^\mu \tilde{F}_{\mu \nu } = J^g_\nu& \\
   &   F_{\mu \nu } = \partial_\mu A_\nu  - \partial_\nu A_\mu &
\end{eqnarray}

where the index $g$ denotes the magnetic current density.

Now the straightforward generalization of systems (1--4) and (5--7) reads

\begin{eqnarray}
   & \partial^\mu F_{\mu \nu } = J_\nu  - m^2A_\nu & \\
   & \partial^\mu \tilde{F}_{\mu \nu } = J^g_\nu &\\
   & \partial^\mu A_\mu  = 0 &\\
   &  F_{\mu \nu } = \partial_\mu A_\nu  - \partial_\nu A_\mu.
\end{eqnarray}

This is the system of generalized Maxwell equations which would presumably
describe the existence of both magnetic charge and non-zero photon mass.

A few remarks are now in order.
\begin{enumerate}
\item The photon mass term $m^2A_\mu $ in the right-hand side of Maxwell
equations violates the symmetry between the electric and magnetic charges.

\item The gauge invariance is completely lost due to the photon mass
term.  Indeed, it can be seen that the transformation
$ A_\mu  \rightarrow A_\mu  + \partial_\mu f $
is inconsistent with equations (8--11) whatever the function $f$ is.
(Recall that the ordinary Maxwell equations in the Lorentz gauge,
$\partial_\mu A^\mu  = 0$, also do not allow gauge transformation with
the arbitrary function $f$.  However, these equations allow such
transformations for $f$ satisfying the condition $\Box f = 0$.  In
our case, even that restricted gauge invariance is lost.)  Note
that this loss of invariance has occurred already at the stage of the
Proca equations without magnetic charge and so it has nothing to do
with the introduction of magnetic charge.

\item Due to the loss of gauge invariance, the vector potential $A_\mu $
becomes observable quantity on the same footing as the field strength
$F_{\mu \nu }$.  It can be seen that the presence of photon mass term
$m^2 A_{\mu}$ in the right-hand side of the  equation (8) creates a sort of
additional current density, in addition to the usual electric current $
j_{\mu}.$
\item  It is not immediately obvious that the loss of gauge invariance destroys
the consistency of the Dirac monopole theory and the validity of the
quantization condition. For example, the Aharonov-Bohm effect which is also
based on the electromagnetic gauge invariance, has been shown to survive in the
massive electrodynamics despite the absence of gauge invariance there
\cite{bd}.
\end{enumerate}

Our modified Maxwell equations tell us that there arises the additional
magnetic field created by the "potential-current" $m^2 A_{\mu}$.  There
is no way to separate this additional magnetic field from the normal one.
Although in Proca theory (without magnetic charge) this circumstance
does not cause any problems, it becomes the main source of trouble once
magnetic charges are added to the massive electrodynamics, as we shall
see shortly.

After these general remarks, let us see if our system of "Maxwell + photon mass
+
magnetic charge" equations (8--11) is consistent or not.  Let us try to
find a static monopole-like solution of that system.  For this purpose, we
assume the absence of electric fields, charges and currents (${\bf E}=0$,
$A_0 = 0$, $\rho = 0$, ${\bf j}= 0$) as well as the absence of magnetic
current (${\bf j}_g=0$).

We then are left essentially with four equations
\begin{eqnarray}
\label{main}
   & \nabla\cdot {\bf H} = \rho_g&\\
   & \nabla\times {\bf H} = -m^2{\bf A}&\\
    &{\bf H} = \nabla\times {\bf A} &\\
    & \nabla\cdot {\bf A} = 0.&
\end{eqnarray}
The first equation has the familiar Dirac monopole solution:
\begin{eqnarray*}
   & {\bf H}^D = \frac{g}{4\pi r^3}{\bf r},&\\
   &A^D_r =\ A^D_\theta = 0,& \\   & A^D_\varphi = \frac{g}{4\pi r} \tan
{\frac{\theta}{2}}.&
\end{eqnarray*}
As is well known, this solution involves a singularity in vector potential
along the line $\theta = \pi$ ("a string").  Yet this singularity has been
shown  to be only a nuisance without any physical significance.
Now if we plug this Dirac solution (or, more exactly, the Coulomb magnetic
field)
into the second equation, we immediately run into trouble, because clearly
$\nabla\times {\bf H}^D = 0$, instead of being equal to $(-m^2{\bf A})$.  Let
us try
to find a better solution by adding something to the Dirac solution.  In this
way we write:
\begin{equation}
        {\bf H} = {\bf H}^D + {\bf H}',\; \; {\bf A} =  {\bf A}^D + {\bf A}',
\end{equation}
where the rotor and the divergence of the additional field ${\bf H}'$ must
satisfy
\begin{equation}
\label{eq:8}
      \nabla\cdot{\bf H}' = 0
\end{equation}
\begin{equation}
\label{eq:9}
      \nabla\times {\bf H}' = -m^2 ({\bf A}^D + {\bf A}'),
\end{equation}
while the divergence of the potential ${\bf A}'$ must vanish:
\begin{equation}
\label{eq:10}
      \nabla\cdot {\bf A}' = 0,
\end{equation}
because
      $\nabla\cdot{\bf A}^D = 0 $  and   $ \nabla\cdot({\bf A}^D + {\bf A}') =
0.$
Finally,
\begin{equation}
\label{eq:11}
      \nabla\times {\bf A}' = {\bf H}'.
\end{equation}

Now, we have the complete system of equations (\ref{eq:8}) through
(\ref{eq:11}) for
the rotors and divergences of both ${\bf H}'$ and ${\bf A}'$.

Let us now find its solution.  Taking the rotor of both
sides of Eqn.(\ref{eq:11}) and using Eqns. (\ref{eq:9}) and (\ref{eq:10}) we
get the second-order
equation for ${\bf A}'$ only:
\begin{eqnarray}
\label{A}
   & (\triangle -m^2){\bf A}'  = m^2{\bf A}^D.&
\end{eqnarray}

In cartesian coordinates we get three decoupled scalar equations instead
of one vector equation (21); after that
 we can use the scalar Green's function to obtain the solution of  Eq.(21) for
${\bf A}'$:

\begin{equation}
\label{eq:star}
{\bf  A'}({\bf r}) =\frac{m^2}{4\pi}\int d^3{\bf r'}{\bf A}^D({\bf
r'})\frac{exp\left(-m|{\bf r} - {\bf r'}|\right)}{|{\bf r} - {\bf r'}|}.
\end{equation}
It can be shown that this solution is indeed transverse:  $\nabla\cdot {\bf
A}'({\bf r})=0$. Also, it can be demonstrated that the potential ${\bf A}'$,
unlike the Dirac potential ${\bf A}^D$, is {\em free from singularities}.
The proof will be published elsewhere.

Let us now find the structural form of the magnetic field  ${\bf H}' ({\bf
r})$.

{}From Eqn.~(\ref{eq:star}) which gives us the integral expression for the
potential ${\bf A}'$,
we can obtain the formula for $\vec{H}'$ by taking rotor:
\begin{eqnarray*}
   {\bf H}'({\bf r}) = \nabla\times {\bf A}' ({\bf r})= \frac{m^2}{4 \pi} \int
d^3 {\bf r}' \frac{e^{-mR}}{R^3} ( 1 + mR) \left( {\bf A}^D \times {\bf R}
\right);  \; \; \; \; {\bf R}={\bf r}-{\bf r}'.
\end{eqnarray*}
Now, because ${\bf H}'({\bf r})$ is a vector (not a pseudovector!) depending on
the two vectors only, ${\bf n}$
and ${\bf r}$, it has the following most general form:
\begin{equation}
\label{eq:mandy}
   {\bf H}'({\bf r}) = h(r, {\bf n}{\bf r}){\bf r} + g(r, {\bf n}{\bf r}){\bf
n}.
\end{equation}
In principle, the functions $h$ and $g$ can be computed explicitely, but we
will not need it.
A question arises: can ${\bf H}'({\bf r}) $ be spherically symmetric, that is,
can it take the
form
$
    {\bf H}'({\bf r}) = h(r){\bf r} \; \;?
$
The answer is no, because in that case we would have
$
    \nabla \times {\bf H}'({\bf r}) = 0
$
everywhere, which is inconsistent with our initial Maxwell equations, see Eq.~
(18).

Having considered the classical theory of massive electrodynamics with magnetic
charge, we can now turn to quantum mechanics. Since 1931, the Dirac
quantization condition has been derived in many ways differing by their initial
assumptions. Obviously those methods using gauge invariance (such as original
Dirac derivation or the Wu-Yang formulation)are not applicable in our case.
Other methods depend on rotational invariance and we can try to generalize them
to include massive electrodynamics, too. Before doing so, let us recall very
briefly the essence of the standard arguments \cite{lwp,h,p,g}.

Consider an electron placed in the field of a magnetic charge $g$. The angular
momentum operator for the electron is given by
$$
{\bf L}={\bf r} \times (-i\nabla + e{\bf A}) + eg \frac{\bf r}{r}, \; \; \; \;
e>0.
$$
Despite the stange-looking second term, this operator can be shown to obey all
the standard requirements of a {\em bona fide \/} angular momentum: see
commutation relations, Eqs.~ (25--27) below. Morover, $L_i$ commute with the
Hamiltonian
\begin{equation}
\label{eq:v}
   H = -\frac{1}{2m}{ (\nabla + ie {\bf A} )}^2 + V(r).
\end{equation}
Next, requiring that the projection of ${\bf L}$ onto the ${\bf r}$ axis, ${\bf
L}{\bf r}/{r}=eg$, should be quantized in the standard quantum-mechanical way,
we obtain
$$
eg= 0, \; \pm \frac{1}{2}, \; \pm 1 \ldots,
$$
which is the Dirac quantization condition. (Of course, such a quick derivation
is not very strict, but it can be elaborated to become quite rigorous).

Now, we would like to generalize this result to the case of massive
electrodynamics. Unfortunately, this turns out to be impossible:
we will show  that the angular momentum operator cannot be defined for the
system of charge
plus monopole within massive electrodynamics. More exactly, the following
theorem holds.

\underline{Theorem}

There are no such operators $L_i$ that the following standard properties are
satisfied:
\begin{equation}
\label{eq:i}
  \left[ L_i, L_j\right] = i \varepsilon_{ijk} L_k
\end{equation}
\begin{equation}
\label{eq:ii}
  \left[ L_i, r_j \right] = i \varepsilon_{ijk} r_k
\end{equation}
\begin{equation}
\label{eq:iii}
  \left[ L_i, D_j \right] = i \varepsilon_{ijk} D_k
\end{equation}

where  ${\bf D} = -i\nabla +   e {\bf A} $
is the kinetic momentum operator,
  $ {\bf D} = m \dot{{\bf r}}$.

Note that the conditions of this theorem are not too restrictive: for example,
we do {\em not} require that the hamiltonian be rotationally invariant (i.e., $
[ L_i, H]=0$ is not required). The proof of the theorem will be given
elsewhere.

To summarize, we have shown that the introduction of an arbitrary small photon
mass makes invalid the existing proofs of the consistency of the Dirac monopole
theory. More exactly, the massive electrodynamics does not allow any
generalization of the methods in which the Dirac monopole was introduced into
the massless electrodynamics. Not only the original Dirac scheme which arrives
at the quantization condition by using the gauge invariance, single-valuedness
of the wavefunction and "the veto" postulate does not work anymore; but also
the different approach relying on the algebra of angular momentum fails in the
case of massive electrodynamics. If the magnetic monopole were ever to be
introduced into massive electrodynamics consistently, that would be possible
only due to some radically new mechanism compared with the existing ones.

What is the physical reason for that failure?

The whole existence of the Dirac monopole in the massless electrodynamics rests
upon the quantization condition which makes invisible the string attached to
the monopole. The quantization condition can be obtained either with the help
of gauge invariance or the angular momentum quantization. In the massive case,
both these approaches are not applicable anymore, as we have shown. That means
that there is hardly any way  to make the string invisible in  massive
electrodynamics.

One may think that our result contradicts the principle of continuity which
states that any physical consequence of massive electrodynamics should go
smoothly into the corresponding result of the standard electrodynamics when the
photon mass tends to zero. Indeed, at first sight the appearance of the Dirac
monopoles at zero photon mass is an obvious discontinuity as compared with
their absence at an arbitrary small photon mass.

However, this simple argument is not yet sufficient to claim discontinuity. An
analogy with a similar ``discontinuity'' is instructive here: consider the
number of photon degrees of freedom in massive and massless electrodynamics.
The photon with a mass has three polarization states, independent of how small
its mass is. Then, as soon as the photon becomes massless, the longitudinal
polarization abruptly vanishes and we are left with only two (transverse)
polarization states. Does this fact create a discontinuity? No. To find out if
there is a discontinuity or not, we have  to study the behavior of a more
physical quantity, such as the probabilities to emit or absorb a longitudinal
photon, rather than merely counting the number of degrees of freedom.

The analysis done by Schr\"{o}dinger and others \cite{bs,gn} shows that if one
considers the interaction of longitudinal photons with matter, this interaction
vanishes as photon mass tends to zero. Thus the longitudinal photons decouple
in the limit $ m_{\gamma} \rightarrow 0 $ so that the continuity is restored.

Coming back to our case with monopoles, one should carry out a similar program
to make sure the continuity is not violated. For example, one can consider the
process of electron scattering off
 the system ``string plus monopole'' in the massive electrodynamics. Then, one
should take the limit  $ m_{\gamma} \rightarrow 0 $ for the cross-section of
such scattering and compare it with the corresponding result for the
electron-monopole scattering in the massless electrodynamics.

Based on our result, one might ask whether the continuous failure to discover
the monopoles in the experiment may be considered as an indirect evidence  for
the finite photon mass.

On the other hand, should the Dirac monopole be found in the future, that would
provide a strong evidence in favour of the exact masslessness of the photon---a
thing impossible to obtain by any direct experimental search for the photon
mass. This is because any experiment can only place a limit on the photon mass
but not prove that this mass is exactly zero. If the monopole exists in
reality, that would not only explain the mysterious fact of the electric charge
quantization, but also prove that the mass of the photon is absolutely zero.

In any case, the present work reveals  a new and important relation between the
two fundamental facts: the masslessness (massiveness?) of the photon and the
non-existence (existence?) of the magnetic monopole.

This work was supported in part by the Australian Research Council.

\end{document}